\documentclass[10pt]{elsart3}
\usepackage{graphicx}
\begin{document}

\newcommand{\be}{\begin{eqnarray}}
\newcommand{\ee}{\end{eqnarray}}
\newcommand{\xx}{\begin{eqnarray*}}
\newcommand{\yy}{\end{eqnarray*}}
\newcommand{\nn}{\nonumber}
\newcommand{\Vol}{{\rm Vol}}
\newcommand{\sign}{{\rm sign}}
\newcommand{\rt}[1]{c_{#1}}
\newcommand{\fa}{f_{\alpha}}
\newcommand{\fb}{f_{\beta}}
\newcommand{\Da}{\Delta_{\alpha}}
\newcommand{\Db}{\Delta_{\beta}}
\newcommand{\D}[1]{\Delta_{#1}}
\newcommand{\G}[1]{\tilde G(#1)}
\newcommand{\Sm}{M}
\newcommand{\mm}{\tilde m}
\newcommand{\rf}[1]{${\rm Eq}.\;(\ref{#1})$}
\newcommand{\rfa}[1]{appendix \ref{#1}}
\newcommand{\rfb}[1]{$\left [{\rm Eq}.\;(\ref{#1})\right ]$}

\begin{frontmatter}
\title{Magnetic oscillations and frequency mixing in a two-band
conductor}
\author[A1]{Jean-Yves Fortin\corauthref{cor}},
\corauth[cor]{Corresponding author, phone number: (33) 3 90 24 06 55}
\ead{ e-mail: fortin@lpt1.u-strasbg.fr}
\author[A2]{Emmanuel Perez} and
\ead{e-mail: backtempus@yahoo.fr}
\author[A2]{Alain Audouard}
\ead{e-mail: audouard@insa-tlse.fr}
\address[A1]{CNRS, UMR 7085, Laboratoire de Physique Th\'eorique,
Universit\'e Louis Pasteur, 3 rue de l'Universit\'e,
67084 Strasbourg, France}
\address[A2]{Laboratoire National des Champs Magn\'etiques Puls\'es,
143 Ave de Rangeuil, 31432 Toulouse, France}

\begin{abstract}
Exact analytical results of the de Haas-van Alphen (dHvA) effect
in an idealized
two-band Fermi liquid with parabolic dispersion are presented.
We consider a Fermi surface consisting in two electron bands with different
band edges and band masses. Magnetic breakthrough (MB) between the bands
is negligible.
Analytical expressions of the dHvA Fourier amplitudes are
derived in the case where the total number of electron is
fixed (Canonical Ensemble, CE). As already reported in the literature,
the oscillations of the chemical potential yield frequency mixing
and Lifshitz-Kosevich (LK) theory, which is valid in the Grand Canonical
Ensemble (GCE), does not apply at very low temperature.
We show that the corresponding Fourier amplitudes
depend on the commensurability between the two
effective masses and also the two fundamental frequencies.
\linebreak

{\it Abstract Identification Number: 053.}

\end{abstract}

\end{frontmatter}

Experimental data on dHvA quantum oscillations in organic conductors
show evidence of a deviation with LK~\cite{LK} and Falikov-Stachowiak (FS)
~\cite {FS} theories for low dimensional systems. We observe in
the Fourier spectrum of the magnetization in the transfer
salt ${\rm\kappa-(BEDT-TTF)_2Cu(NCS)_2}$~\cite{uji.97}
frequencies that are not related to
a classical orbit on the Fermi surface. For these ``forbidden'' frequencies,
the electron orbits locally in the opposite direction
imposed by the magnetic field. This material is made of 2 disconnected
Fermi surfaces, one pocket hole and an open one dimensional surface.
From LK and FS theories, we should obtain a small frequency
$f_{\alpha}$ corresponding to the pocket, and, by MB
effect between the pocket and the open surface, a larger frequency
$f_{\beta}>f_{\alpha}$ at higher
field ($>$20T at 1K), plus their harmonics and classical combinations that
are permitted like $f_{\alpha}+f_{\beta}$, $2f_{\beta}-f_{\alpha}$ etc...
due to MB and Bragg reflections throughout the Brillouin zone.
The ``forbidden'' frequencies seen experimentally correspond to
$f_{\beta}-f_{\alpha}$, $2f_{\beta}-2f_{\alpha}$.
In Shubnikov-de Haas (SdH) transport experiments, they may be explained
by quantum interference or Stark
effect, like in $(BEDO-TTF)_5[CsHg(SCN)_4]_2$~\cite{lyub.02}.
In dHvA experiments, the explanation seems rather different because
Stark effect does not apply.
There is an argument~\cite{alx.96,harrison.96,fortin.98} showing that
fluctuations of the chemical potential are important in these low dimensional
systems, especially at very low temperature ($<$1K) in the high field range.
This implies that LK and FS frequencies and Fourier amplitudes
should be different in the CE, where the electron density $n_e$ is
fixed, and that the classical way of defining a frequency as area of a
classical orbit is incorrect in this ensemble.
 However, since the LK and FS are correct in the GCE, thermodynamical
relations connect the 2 Ensembles. Determinating the
chemical potential is not simple in the finite temperature case and a
complete calculation is necessary. However we can obtain analytical results
at T=0 in some cases~\cite{fortin.98}.

 In order to reproduce the main features of transfer salt Fermi surface,
we condider a two-band model with gaps $\Delta_{\alpha=0,1}$ and
effective masses $m_{\alpha=0,1}^*$ in an external magnetic field $B$
(we neglect magnetic interaction). This model has
been considered by other authors mostly in a numerical
way~\cite{alx.96,alx.01}.  However there is no MB unlike BEDT systems.
 The Landau levels (LL) of each band $\alpha=0,1$ have energies:

\be\label{eq1}
\epsilon_{\alpha}(n)=\Delta_{\alpha}+(h^2/2\pi m_{\alpha}^*)(eB/h)(n+1/2).
\ee

We will define energies in unit of $2\pi\hbar^2/m_0^*$, and $b=eB/h$ is the
dimensionless
field per unit cell area. In absence of field, the electron densities
$n_{\alpha}$ of each band satisfy the conservation equation
$\sum_{\alpha}n_{\alpha}=n_e$.
If $S_{\alpha}$ is the area of the $\alpha$-band
Fermi surface, then $n_{\alpha}=S_{\alpha}/4\pi^2$. In term of the
Fermi energy $E_f$, we have $S_{\alpha}=2\pi m_{\alpha}^*/\hbar^2
(E_f-\Delta_{\alpha})$.
In the following, we will use the dimensionless variable $x=n_e/b$, and
the ratios $\rt{\alpha}=m_0^*/m_{\alpha}^*$.
In the new units, the LL are $\epsilon_{\alpha}(n)=
\Delta_{\alpha}+\rt{\alpha}b(n+1/2)$.
It is also convenient to use the dimensionless fundamental frequencies
$\fa=S_{\alpha}/4\pi^2n_e$, with $f_0+f_1=1$. Defining
$f_b=(\Delta_0-\Delta_1)/\rt{1} n_e$, we obtain

\be\nn
f_0=\frac{\rt{1}}{1+\rt{1}}(1-f_b),\;\;{\rm and}\;\;f_1=\frac{1}{1+\rt{1}}
(1+\rt{1} f_b)
\ee

From the LK theory, the Grand Potential $\Omega$ is the sum of
each band contribution $\Omega_{\alpha}$:

\be\label{eq2}
\Omega_{\alpha}&=&-\frac{1}{2\rt{\alpha}}(\mu-\Delta_{\alpha})^2+
\frac{b^2\rt{\alpha}}{2}\left [
\frac{1}{12} +\right .
\\ \nn
& &\left .\sum_{p\ge 1}\frac{(-1)^p}{\pi^2p^2}R_{\alpha,p}(T)\cos\left (2\pi p
\frac{\mu-\Delta_{\alpha}}{\rt{\alpha}b}\right )\right ]
\ee

$R_{\alpha,p}(T)=p\lambda_{\alpha}
/\sinh (p\lambda_{\alpha})$ is the temperature reduction factor,
and $\lambda_{\alpha}=2\pi^2T/\rt{\alpha}b$.
The oscillating part of the magnetization
$M=-\partial \Omega/\partial B$
contains cosine functions of arguments $2\pi p F_{\alpha}/B=2\pi p
\fa x$ with $p$ integer, and therefore the Fourier
spectrum is made of the individual frequencies $F_{\alpha}$ plus their
harmonics. There is no mixing.
In the $CE$, we will show that the oscillating part of the
magnetization $M=-\partial F/\partial B$,
where $F=\Omega+n_e\mu$ is the free energy, contains the
combinations $kf_0+lf_1$.
The chemical potential at finite field is the implicit solution
of the conservation equation $\partial\Omega/\partial\mu=-n_e$.
Replacing the solution $\mu$
into \rf{eq2} for $\Omega$, we can obtain the free energy,
hence $M$.

At zero temperature, we can solve the chemical potential equation exactly
and obtain the exact total energy $E$ as a function of $x$. $E$ can be
expressed as $E_0+b^2G(x)/2$,
with $G(x)$ an oscillatory function, and $E_0$ the zero field energy :

\be\label{eq9}\nn
E_0&=&\frac{\rt{1}}{2(1+\rt{1})}n_e^2+\frac{n_e}{1+\rt{1}}
(\rt{1}\D{0}+\D{1})\\
&-&\frac{(\D{0}-\D{1})^2}{2(1+\rt{1})}
\ee

We used a combinatorial analysis to compute $G$. We had to
find the $[x]+1$ lowest LL between the 2 bands, with the first $[x]$
LL completely filled and the last partially.
After tedious computations, we arrive at the following expression:

\be\label{eq22}
& &G(x)=\frac{1}{12}\frac{1+4\rt{1}+\rt{1}^2}{1+\rt{1}}
\\ \nn
&-&\frac{\rt{1}}{1+\rt{1}}\sum_{l=1}^{\infty}
\frac{1}{\pi^2l^2}\cos\left (2\pi l(f_0+f_1)x\right )
\\ \nn
&+&\sum_{l\ge 1}\frac{(-1)^{l}}
{2\pi^4l^4}\frac{(1+\rt{1})^3}{\rt{1}^2}
\left (1-\cos\frac{2\pi l\rt{1}}{1+\rt{1}}\right )
\cos\left (2\pi lf_0 x\right )
\\ \nn
&+&2\sum_{l\ge 1}\sum_{l'\ne l}\frac{(-1)^{l'-l}}
{4\pi^4(l'-l)^2}\frac{(1+\rt{1})^3}{\left (l+l'\rt{1}
\right )^2}
\\ \nn
&\times&\left (1-\cos\frac{2\pi(l'-l)\rt{1}}{1+\rt{1}}\right )
\cos\left (2\pi(l'f_0+lf_1)x\right ).
\ee

\rf{eq22} is the main
result for the two-band problem at zero temperature. The
oscillating part of the magnetization is simply
$m_{osc}=n_eG'(x)/2=\sum_FA(F)\sin(2\pi Fx)$.
In the double sum, it can happen that some terms
diverge when $l+l'\rt{1}$ vanishes in the denominator.
But in this case, the quantity $(l-l')\rt{1}/(1+\rt{1})=l$ is integer
and the ``interference'' term $1-\cos(2\pi l)$ in the numerator
vanishes at the same time.
These divergences are regularized by taking the finite limit of
the quantity $(1-\cos(\epsilon))/\epsilon^2$ when $\epsilon\rightarrow 0$.

\rf{eq22} gives the amplitudes for
the $k^{th}$ harmonics of the combinations $F=l_0f_0+l_1f_1$.
$l_1$ is a positive integer
and $l_0$ can be negative. These include ``borbidden'' frequencies like
$f_1-f_0$. The amplitudes depend on whether $f_0/f_1$ is equal to
an irreducible ratio of 2 integers $p_1/p_0$. In this case, we can find
many combinations $l'f_0+lf_1$ that are equal to $kF$. These occur
when $l=kl_1+mp_1$ and $l'=kl_0-mp_0$, with $m$ positive integer.
We then obtain

\be\nn
A(kF)&=&\frac{kF(1+\rt{1})^3}{2\pi^3}
\\ \nn
&\times& \sum_{m\ge 0}
\frac{(-1)^{k(l_0-l_1)-m(p_0+p_1)+1}}{(k(l_1+\rt{1}l_0)+m(p_1-\rt{1}p_0))^2}
\\ \label{eq33}
&\times &\frac{1-\cos \frac{2\pi\rt{1}\left (k(l_0-l_1)
-m(p_0+p_1)\right )}{1+\rt{1}}}{(k(l_0-l_1)-m(p_0+p_1))^2}.
\ee

In the case where $f_0/f_1$ is not rational, only the term $m=0$ has
to be taken.
We will only consider in the rest of the paper the case when $f_0/f_1$
is irrational, for simplication, and therefore only the first term $m=0$
on the right hand side of \rf{eq33} contribute.
\rf{eq33} holds for any frequency except
for the special case $F=k(f_0+f_1)$:

\be\label{eq26}
A(k(f_0+f_1))=\frac{\rt{1}}{1+\rt{1}}\frac{1}{\pi k}.
\ee

We obtain a simple result:
the $k^{th}$ harmonics amplitude of $f_0+f_1$ is the same as it would be
in GCE case for
a single band with fundamental frequency $f_0+f_1$ (except for
a $(-1)^k$ sign). The amplitudes for harmonics of $f_{\alpha}$ are:

\be\nn
%\label{eq31}
A(kf_{\alpha})=\frac{(-1)^{k+1}(1+\rt{1})^3f_{\alpha}}{2\pi^3k^3\rt{\alpha}^2}
\left (1-\cos\frac{2\pi k\rt{1}}{1+\rt{1}}\right ).
\ee

The difference with the GCE is that harmonics amplitudes of any frequency but
$f_0+f_1$ fall like $1/k^3$
instead $1/k$. We conclude that the jumps in magnetization only
come from the contribution of $f_0+f_1$, since its $k^{th}$ harmonics
amplitude decreases like $1/k$.
The amplitude of the ``forbidden'' frequency $f_1-f_0$ is:

\be\nn
A(f_1-f_0)=-\frac{(1+\rt{1})^3}{(1-\rt{1})^2}\frac{f_1-f_0}{(2\pi)^3}
\left (1-\cos\frac{4\pi\rt{1}}{1+\rt{1}}\right )
\ee

\begin{figure}
\begin{center}
\includegraphics*[scale=0.45]{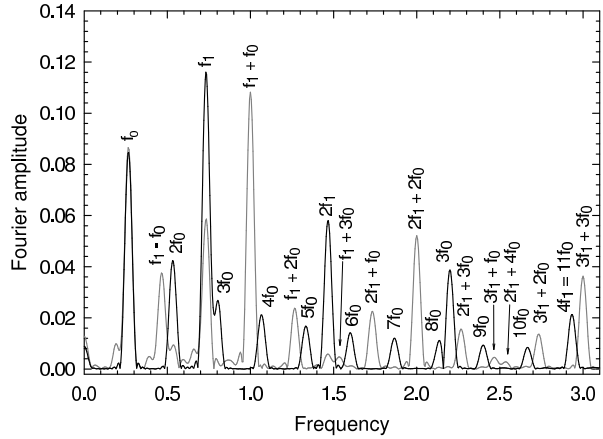}
\caption{Comparison between the absolute amplitudes in the CE
(grey line) and GCE (black line) for $c_1=m^*_0/m^*_1=1/2$,
$\Delta_0=0.1$, $\Delta_1=0$, $n_e=1$, $f_0=4/15$ and $f_1=11/15$.
}
\end{center}
\end{figure}

Figure 1 shows the numerical comparison between GCE and CE amplitudes.
Analytical approximation based on
thermodynamical relations~\cite{alx.01} shows that the magnetization
amplitudes $A(F)$ for combined frequencies $F=l_0f_0+l_1f_1$ at
finite temperature
are proportional to $R_{0,l_0}(T)R_{1,l_1}(T)F/l_0l_1$. Basically
this is the product of GCE amplitudes for the individual frequencies
$l_0f_0$ and $l_1f_1$. For $F=kf_{\alpha}$, the authors found $A(F)\propto
R_{\alpha,k}f_{\alpha}/k$.
At zero temperature the $k^{th}$ harmonics amplitudes of these frequencies
decrease to $1/k$ in any case, which is not the exact
case since we showed the
dependence in $k$ is $1/k^3$.
Moreover, the effect of commensurability between fundamental
frequencies is not included. The finite low temperature dependence
of the amplitudes is therefore not understood yet.
However their approximation may work beyond
a crossover temperature below which corrective terms are
needed. Indeed the chemical potential fluctuations decrease with increasing
temperature and we expect GCE and CE temperature reduction factors be
equivalent for frequencies $kf_{\alpha}$.

In conclusion, we computed the exact Fourier magnetization for the 2 band model
at zero temperature. The expressions found are different from LK theory
except for the harmonics of $f_0+f_1$.
The $k^{th}$ harmonics amplitude of other frequencies decreases like $1/k^3$,
and includes an
interference term depending on the mass ratio. These results may be
useful to check possible low temperature theories.

\end{document}